\documentclass[11pt,a4paper]{article}
\pdfoutput=1
\usepackage{jcappub}

\usepackage{graphicx}
\usepackage{amsmath,amssymb,amsfonts}
\usepackage{verbatim}  
\usepackage{subfigure} 
\usepackage{acronym}
\usepackage{hyperref}
\usepackage{mathrsfs}
\usepackage{multirow}
 \usepackage{slashed}

\usepackage{amsmath}
\usepackage{amsfonts}
\usepackage{amssymb}

\newcommand{\beq}{\begin{equation}}
\newcommand{\eeq}{\end{equation}}
\newcommand{\bea}{\begin{eqnarray}}
\newcommand{\eea}{\end{eqnarray}}

\let\oldmarginpar\marginpar
\renewcommand\marginpar[1]{\-\oldmarginpar[\raggedleft\scriptsize\sf
#1]{\raggedright\scriptsize\sf #1}} 

\begin{document}

\arxivnumber{OUTP-13-06P}

\title{\boldmath On the importance of loop-induced spin-independent interactions for dark matter direct detection\unboldmath}

\author{Ulrich Haisch}
\author{and Felix Kahlhoefer}
\emailAdd{u.haisch1@physics.ox.ac.uk}
\emailAdd{felix.kahlhoefer@physics.ox.ac.uk}
\affiliation{Rudolf Peierls Centre for Theoretical Physics, University of Oxford, 1 Keble Road, \\ Oxford OX1 3NP, United Kingdom}

\date{\today}

\abstract{
The latest results from LHC searches for jets in association with missing transverse energy place strong bounds on the scattering cross section of dark matter. For the case of spin-dependent or momentum suppressed interactions these limits seem to be superior to the bounds from direct detection experiments. In this article, we show that loop contributions can significantly alter this conclusion and boost direct detection bounds, whenever they induce spin-independent interactions. This effect is most striking for tensor and pseudotensor interactions, which induce magnetic and electric dipole moments at loop level. For axialvector and anapole interactions a relevant contribution to direct detection signals arises from loop-induced Yukawa-like couplings  between dark matter and quarks. We furthermore compare the resulting bounds to additional constraints on these effective operators arising from indirect searches and relic density requirements.
}


\maketitle

\section{Introduction} 
\label{sec:intro}

``What is dark matter made of?" is one of the big, unsolved questions of modern particle physics. Three kinds of search strategies promise to shed light on this issue: direct detection experiments attempt to observe the interaction of dark matter (DM) particles in shielded underground detectors. Indirect detection experiments look for the products of DM annihilations with satellites, balloons and ground based telescopes. Finally, accelerator experiments aim to actually produce DM by colliding Standard Model (SM) particles at high energies.

These searches are complementary in the sense that each of them is best suited to probe certain mass regions and types of DM interactions. For example,  light DM is difficult to explore with direct detection, because a DM particle with a mass of a few GeV typically cannot produce nuclear recoil signals above the detector threshold~\cite{Altmann:2001ax,Ahmed:2009zw,Ahmed:2010wy,Bernabei:2010mq,Aalseth:2011wp,Felizardo:2011uw,Archambault:2012pm,Behnke:2012ys,KIMS,Aprile:2012nq,Aprile:2013uj}. No such limitation exists at colliders, since in the low-mass window  monojet and monophoton cross sections, as well as the kinematic distributions, become essentially independent of the DM mass~\cite{Chatrchyan:2012me,ATLAS:2012ky}. On the other hand, collider searches cease to be constraining for DM particles with mass of a TeV or more, making direct and indirect searches the most promising strategies for these parameters.

Similarly, direct detection experiments give the strongest constraints on spin-in\-de\-pen\-dent~(SI) interactions, because the coherent scattering leads to an enhancement proportional to the square of the nucleus mass. For interactions that are either spin-dependent~(SD) or momentum suppressed  direct detection constraints on the scattering cross section are weakened by many orders of magnitude. For collider searches, on the other hand, the cross section for the process $pp \to \chi \chi + {\rm jet(s)}/\gamma$ is  largely independent of the Lorentz structure for four-fermion operators with the same couplings to light quarks.

Recent monojet searches at the LHC seem to suggest that direct detection searches are generally inferior to collider searches in constraining the cross section for SD and momentum suppressed  interactions of DM particles with mass below the TeV scale. As we will demonstrate in this article, this is not necessarily the case, because a rigorous classification of DM-nucleon interactions into SI and SD (or momentum suppressed) is not possible in general, as such a distinction is unstable under radiative corrections. We will illustrate this feature by considering two types of effective operators which lead to either SD or momentum suppressed  interactions at tree level, but induce SI interactions once loop contributions are included.

For operators with tensor or pseudotensor structure  we show that such couplings lead to a magnetic or electric dipole moment of the DM particle, respectively, corresponding to DM-photon interactions arising from the virtual exchange of heavy quarks. While the former interactions are SD or momentum suppressed and therefore pose a challenge for current (and future) direct detection experiments, the latter operators lead to an SI part in the differential cross section that is enhanced in the infrared due to the photon pole. This feature makes DM with dipole-type interactions relatively easy to detect in low-energy scattering. We find that by including loop contributions, the existing bounds on the new-physics scale that suppresses the four-fermion operators are improved by a factor of around 3 (120) in the case of tensor (pseudotensor) interactions. As a result, present direct detection experiments allow to indirectly probe scales of up to $3.7 \, {\rm TeV}$ ($159 \, {\rm TeV}$) associated to these specific operators.

For the case of the axialvector and anapole operators  we consider loop diagrams involving two operator insertions and show that such corrections generally give rise to SI interactions between DM and quarks. The induced effective interactions are proportional to the quark mass and the mass of the DM particle. For the axialvector operator  the resulting constraints are weaker than those obtained at tree level, but for the anapole operator the two different contributions turn out to be comparable in parts of the parameter space.

While prior studies (see {\it e.g.}~\cite{Cirelli:2005uq,Essig:2007az,Kopp:2009et, Freytsis:2010ne, Fox:2011fx, Hill:2011be, Weiner:2012cb, Frandsen:2012db, Haisch:2012kf,Fox:2012ru}) have stressed the importance of radiative corrections in DM physics, we feel that these ideas have not been fully embraced by the community. We hope that the results presented in this paper convince the reader that the time is ripe for more detailed studies of loop effects in the DM context. 

Our work is organised as follows. After listing the effective operators relevant to our analysis in Sec.~\ref{sec:operators}, we turn to the tensor and pseudotensor operators in Sec.~\ref{sec:tensor} and calculate the loop-induced magnetic and electric dipole moments. In this section we also determine the bounds on the new-physics scale that arise from direct detection experiments and compare them to other existing limits. We then examine the axialvector and anapole operators in Sec.~\ref{sec:axial} and calculate the SI cross section arising from loop-level processes. We also discuss the impact of our findings in the context of relic density constraints and existing anomalies in the DM sector. An appendix  contains details concerning technical aspects of our calculations.

\section{Effective Operators} 
\label{sec:operators}

In the following our focus will be on the interactions between a fermionic DM particle $\chi$ and SM quarks as well as photons. We assume that $\chi$ is a Dirac fermion.\footnote{For the case of Majorana fermions the tensor and pseudotensor  operators vanish, so we do not discuss this case here. We note, however, that the discussion of axialvector and anapole operators presented in~Sec.~\ref{sec:axial} applies to Majorana fermions as well. In this case, the loop-level bounds shown in Fig.~\ref{fig:Anapolebound} would be stronger by a factor of 4, while the tree-level bounds remain unchanged.}  In order to keep our discussion as general as possible, we will employ an effective field theory obtained by integrating out heavy degrees of freedom above a certain high-energy scale $M_\ast$. The resulting effective operators can be classified by their behaviour in the non-relativistic limit. The only four-fermion operators leading to unsuppressed SD interactions are the axialvector operator $\mathcal{O}_{AX}$ and the tensor operator $\mathcal{O}_T$~\cite{Kopp:2009qt}. We write these operators as
\beq \label{eq:SD}
\mathcal{O}_{AX} = \frac{1}{M_\ast^2} \left ( \bar \chi \hspace{0.25mm} \gamma_\mu \gamma_5 \hspace{0.25mm} \chi \right ) \left ( \bar q \hspace{0.25mm} \gamma^\mu \gamma_5 \hspace{0.25mm} q \right )\,, \qquad 
\mathcal{O}_T = \frac{1}{M_\ast^2} \left ( \bar \chi \hspace{0.25mm} \sigma_{\mu \nu} \hspace{0.25mm} \chi \right ) \left ( \bar q \hspace{0.25mm} \sigma^{\mu \nu} \hspace{0.25mm} q \right ) \,,
\eeq
where the quark field $q$ can be of arbitrary flavour and $\sigma_{\mu \nu} = i/2 \hspace{0.5mm} (\gamma_\mu \gamma_\nu - \gamma_\nu \gamma_\mu)$. 

Various other four-fermion operators give interactions which are suppressed in the non-relativistic limit, either by a factor $q^2/m_N^2 \ll 1$, where $q$ is the momentum transfer in the low-energy scattering and $m_N$ is the mass of the target nucleus, or by a factor $v^2 \ll 1$, where $v$ is the velocity of the DM particle. We shall only consider two of these operators here, namely the anapole operator $\mathcal{O}_{AN}$~\cite{Fitzpatrick:2010br,Ho:2012bg} and the pseudotensor operator $\mathcal{O}_{PT}$. We define them as
\beq \label{eq:SDsup}
\mathcal{O}_{AN} = \frac{1}{M_\ast^2} \left ( \bar \chi \hspace{0.25mm} \gamma_\mu \gamma_5 \hspace{0.25mm} \chi \right ) \left ( \bar q \hspace{0.25mm} \gamma^\mu \hspace{0.25mm} q \right ) \,, \qquad 
\mathcal{O}_{PT} = \frac{i}{M_\ast^2} \left ( \bar \chi \hspace{0.25mm} \sigma_{\mu \nu} \gamma_5 \hspace{0.25mm} \chi \right ) \left ( \bar q \hspace{0.25mm} \sigma^{\mu \nu} \hspace{0.25mm} q \right )\,.
\eeq
Unsuppressed SI interactions can, on the other hand, be obtained from the scalar operator $\mathcal{O}_S$ and the vector operator $\mathcal{O}_V$~\cite{Beltran:2008xg}, which are of the form
\beq \label{eq:SI}  
\mathcal{O}_{S} = \frac{m_q}{M_\ast^3} \hspace{0.75mm} {\cal C}_S \left ( \bar \chi \hspace{0.25mm} \chi \right ) \left ( \bar q \hspace{0.25mm} q \right ) \,, \qquad 
\mathcal{O}_{V} = \frac{1}{M_\ast^2} \hspace{0.75mm} {\cal C}_V \left ( \bar \chi \hspace{0.25mm} \gamma_\mu \hspace{0.25mm} \chi \right ) \left ( \bar q \hspace{0.25mm} \gamma^\mu \hspace{0.25mm} q \right )\,.
\eeq
Finally, SI scattering can also result from the magnetic and electric dipole-type interactions~\cite{Barger:2010gv,Banks:2010eh} encoded in
\beq \label{eq:dipole}  
\mathcal{O}_M = \frac{1}{M_\ast^2} \hspace{0.75mm} {\cal C}_M \left ( \bar \chi \hspace{0.25mm} \sigma_{\mu \nu} \hspace{0.25mm} \chi \right ) F^{\mu \nu} \,, \qquad 
\mathcal{O}_E = \frac{i}{M_\ast^2} \hspace{0.75mm} {\cal C}_E  \left ( \bar \chi \hspace{0.25mm} \sigma_{\mu \nu} \gamma_5 \hspace{0.25mm} \chi \right ) F^{\mu \nu} \,,
\eeq
with $F^{\mu \nu}$ denoting the regular electromagnetic field strength tensor. The coefficient ${\cal C}_M$~(${\cal C}_E$) can be interpreted as the magnetic (electric) dipole moment of the DM particle in units of $M_\ast^2$. The resulting cross sections for DM-nucleon scattering via photon exchange are {\it enhanced} for small momentum transfer, due to the propagator from the massless mediator.

Even though we employ a low-energy effective description in what follows, we should briefly discuss ultraviolet (UV) completions for the above operators. The operators ${\cal O}_{AX}$ and ${\cal O}_{AN}$ can arise from the $s$-channel exchange of a heavy spin-1 mediator. Additional operators, such as ${\cal O}_{V}$ can be absent either because of a suitable choice of couplings or because the DM vector current vanishes (for example if the DM particle is a Majorana fermion). For alternative models that give rise to ${\cal O}_{AN}$, we refer to \cite{Fitzpatrick:2010br,Ho:2012bg}. The operators ${\cal O}_T$ and ${\cal O}_{PT}$, on the other hand, arise from Fierz rearrangements of effective operators like $(\bar \chi q) (\bar q \chi)$ and $(\bar \chi \gamma_5 q) (\bar q \gamma_5 \chi)$, which result from integrating out a $t$-channel mediator (such as the heavy spin-0 particles present in models of top-flavoured minimal flavour violating DM~\cite{Batell:2011tc}). In this case, other effective four-fermion operators will be present in the low-energy description. To study the interplay of different operators is beyond the scope of this paper. For our purposes it is sufficient to note that there are UV completions which induce the operators ${\cal O}_{AX}$ and ${\cal O}_{AN}$, but not the scalar and vector operators, and UV completions which induce ${\cal O}_{T}$ and ${\cal O}_{PT}$, but not the electric and magnetic dipole operators.

Various bounds on the operators in~(\ref{eq:SD}) to (\ref{eq:dipole}) have been derived in the DM literature~\cite{Beltran:2008xg,Harnik:2008uu,Kopp:2009qt,Agrawal:2010fh,Goodman:2010ku,Goodman:2010qn,Barger:2010gv,Banks:2010eh,Fitzpatrick:2010br,Cheung:2010ua,Zheng:2010js,Fortin:2011hv,Rajaraman:2011wf,Cheung:2011nt,Cheung:2012gi,Fox:2012ee,MarchRussell:2012hi,Barger:2012pf,Chatrchyan:2012me,ATLAS:2012ky,Ho:2012bg,Zhou:2013fl} and we will review the most important ones in the next section.  At this point, it is sufficient to recall that for both SI and SD interactions the DM-nucleon scattering cross section is approximately given by $\sigma \propto m_\text{red}^2 / M_\ast^4$, where $m_\text{red}$ is the reduced mass of the DM-nucleon system (the precise formulas are given in (\ref{eq:SDDD}) and (\ref{eq:SIDD})). The DM-\emph{nucleus} scattering cross section, on the other hand, is typically enhanced for SI interactions compared to SD interactions by a factor of $A^2 = {\cal O} (10^4)$, where $A$ is the mass number of the nucleus. For this reason constraints from direct detection experiments on SD interactions are typically much weaker than constraints on SI interactions (see also~\cite{Kopp:2009qt}). If the interaction is additionally suppressed by powers of $q^2 / m_N^2$ or $v^2$, the constraints are completely negligible. 

These features suggest that LHC searches involving missing transverse energy (MET)  provide the strongest constraints on $\mathcal{O}_T$ and the {\it only} constraint on the operator $\mathcal{O}_{PT}$. This naive expectation, however, turns out to be incorrect. The reason is that each of these operators induces one of the SI operators given in~(\ref{eq:dipole}) at the one-loop level. Since the experimental bounds on the SI operators from direct detection are very constraining, these searches then also give the leading limits on the SD operators over a wide range of the parameter space. 

For the operators $\mathcal{O}_{AX}$ and $\mathcal{O}_{AN}$ we make a similar observation: direct detection cross sections are suppressed at tree-level, but at the one-loop level, the SI operator $\mathcal{O}_{S}$ is induced. Again, we can use the strong experimental bounds on this operator to obtain competitive bounds on the axialvector and anapole operators. We will now discuss each of these cases in turn.

\section{Constraints on tensor and pseudotensor operators}
\label{sec:tensor}

\begin{figure}[!t]
\begin{center}
\includegraphics[height=0.25 \textwidth]{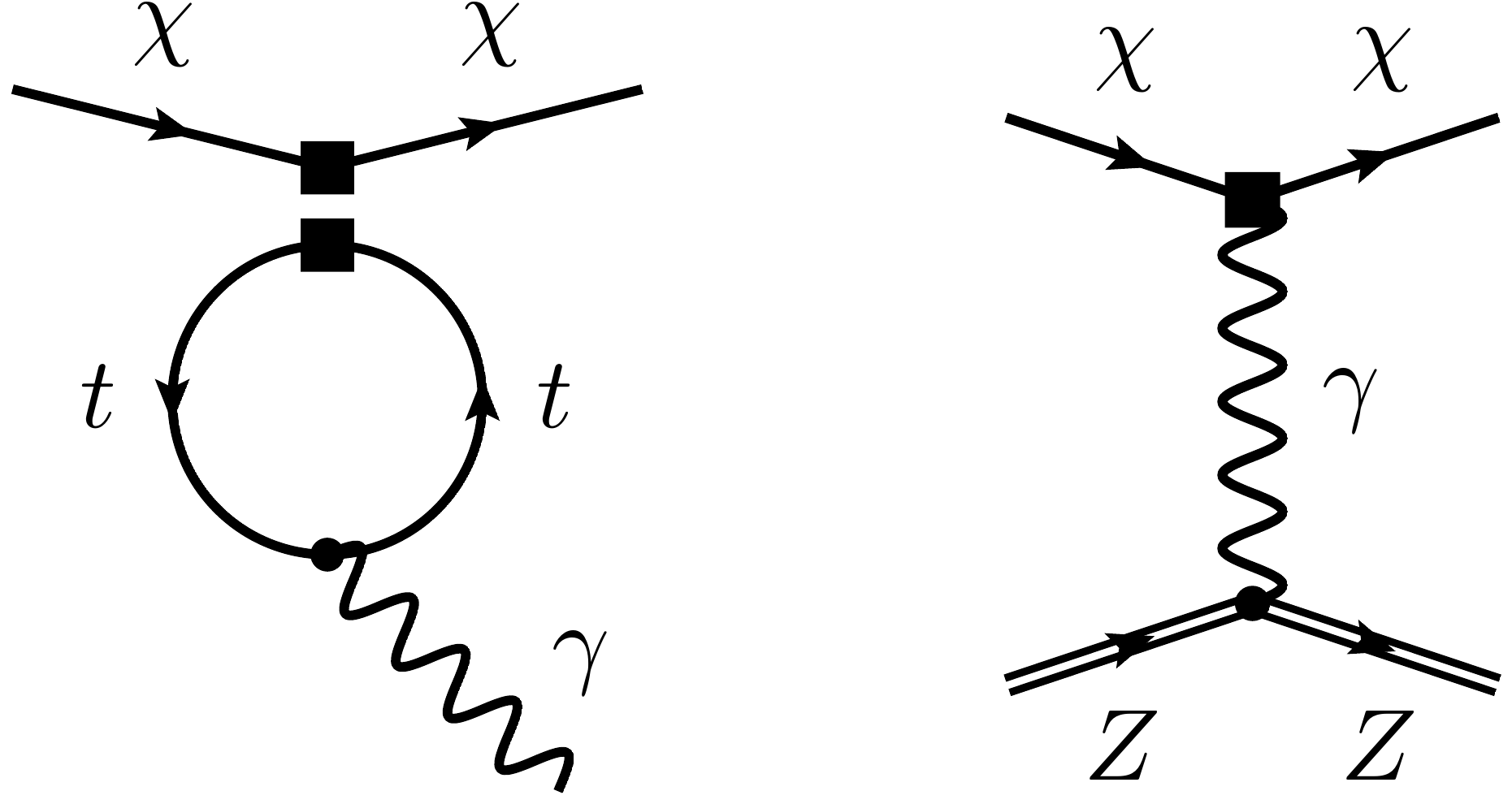}
\vspace{1mm}
\caption{\label{fig:diagram} Left: One-loop diagram with a closed top-quark loop that leads to a dipole-type operator. 
Right: Feynman diagram for elastic DM scattering on a nucleus with electric charge $Z$. The black squares in both graphs denote an operator insertion. }
\end{center}
\end{figure}

We consider the one-loop diagram shown on the left in Fig.~\ref{fig:diagram}. A straightforward calculation (outlined in App.~\ref{app:1}) shows that the insertion of $\mathcal{O}_T$ into this graph induces a contribution to $\mathcal{O}_{M}$ with a magnetic dipole moment (in units $M_\ast^2$) that is approximately given by 
\beq \label{eq:MLL} 
{\cal C}_M \simeq \frac{3  \hspace{0.25mm}  e  \hspace{0.25mm}  Q_t}{4 \pi^2} \hspace{0.5mm} m_t  \hspace{0.25mm} \ln \frac{M_\ast^2}{m_t^2} \,,
\eeq
where $Q_t = 2/3$ and $m_t \simeq 163 \, {\rm GeV}$ are the electric charge and the mass of the top quark and we have assumed that $M_\ast > m_t$. Due to the chiral invariance of QED and QCD the above result also holds for the coefficient ${\cal C}_E$ that results from the mixing of $\mathcal{O}_{PT}$ into $\mathcal{O}_E$. In order to get a more accurate estimate of ${\cal C}_M$, we resum the leading logarithms appearing in~(\ref{eq:MLL}) by employing renormalization group~(RG) techniques (see {\it e.g.}~\cite{Bobeth:2011st}). Numerically this resummation alters the induced magnetic dipole moment by below  $-10\%$ for new-physics scales $M_\ast$ in the range $[1, 100] \, {\rm TeV}$.  Notice that also bottom- and charm-quark loops with an insertion of $\mathcal{O}_{T}$ and $\mathcal{O}_{PT}$ will induce dipole moments. The corresponding expression for ${\cal C}_M$ (${\cal C}_E$) is obtained from (\ref{eq:MLL}) by the simple replacements $Q_t \to Q_{b,c}$ and $m_t \to m_{b,c}$. These findings agree with the discussion of loop-induced DM dipole moments in the context of leptophilic DM~\cite{Kopp:2009et}.

\subsection{Direct detection bounds on dipole moments}
\label{sec:dipolebounds}

Direct detection experiments aim to observe DM by measuring the energy transferred from DM particles to target nuclei in elastic scattering processes. In the case of the dipole-type operators (\ref{eq:dipole}) this scattering proceeds through the exchange of a photon, as shown on the right-hand side of Fig.~\ref{fig:diagram}. In the non-relativistic limit the corresponding differential cross sections are~\cite{Barger:2010gv,Banks:2010eh}
\beq
\begin{split}
\label{eq:crosssection}
\frac{d\sigma_M}{dE_\mathrm{R}} & = \frac{4 \alpha \hspace{0.25mm} {\cal C}_M^2}{M_\ast^4} \, \bigg [ \hspace{0.25mm} Z^2 \left(\frac{1}{E_\mathrm{R}} - \frac{1}{m_\chi v^2} - \frac{1}{2 m_N v^2}\right)\left|F_c(E_\mathrm{R})\right|^2 \\ 
& \hspace{1.75cm} +  \frac{2 \left ( J+1 \right) }{3J}\left(\frac{\mu_N A}{\mu_n}\right)^2 \frac{1}{2 m_N v^2} \left|F_s(E_\mathrm{R})\right|^2 \bigg ] \,, \\[2mm]
\frac{d\sigma_E}{dE_\mathrm{R}} & = \frac{4 \alpha \hspace{0.25mm} {\cal C}^2_E}{M_\ast^4} \frac{Z^2}{v^2 E_\mathrm{R}} \left|F_c(E_\mathrm{R})\right|^2 \,.
\end{split}
\eeq
Here $\alpha \simeq 1/137$ is the fine-structure constant of electromagnetism, $J$ is the spin of the target nucleus, $m_N$ is its mass and $\mu_N / \mu_n$ is its magnetic moment in units of the Bohr magneton $\mu_n = e/(2 \hspace{0.25mm} m_n)$. $Z$ and $A$ are the charge and mass number of the nucleus, respectively. The form factors $F_s (E_{\mathrm{R}})$ and $F_c (E_{\mathrm{R}})$ reflect the distribution of spin and charge in the nucleus as a function of the recoil energy $E_{\mathrm{R}}$, and we adopt the parameterisations introduced in~\cite{Banks:2010eh} to model these functions. Since most collisions have a momentum transfer small compared to the inverse nuclear radius, uncertainties in the form factors have a negligible effect on our final results. We observe that all terms in the expression for $d\sigma_M/dE_\mathrm{R}$ are suppressed compared to $d\sigma_E/dE_\mathrm{R}$ either by a factor of $v^2$, by a factor of $E_\mathrm{R} / m_\chi$ or by a factor of $E_\mathrm{R} / m_N$, which are all typically ${\cal O} (10^{-6})$. Consequently, effective operators that induce an electric dipole moment are in general the most severely bounded.
\begin{table}[t!]
\setlength{\tabcolsep}{5pt}
\renewcommand{\arraystretch}{1.2}
\center
\begin{tabular}{|c|c|c|} 
\hline
$m_\chi$ [GeV] & $M_\ast$ in $\mathcal{O}_T$ [GeV] & $M_\ast$ in $\mathcal{O}_{PT}$ [TeV] \\
\hline \hline 
10 & $1880^{+360}_{-450}$ & $65 .6^{+5.5}_{-5.6}$ \\
\hline
20 & $3360^{+520}_{-600}$ &  $123 .7^{+9.6}_{-9.6}$ \\
\hline
50 & $3740^{+560}_{-640}$ & $158 .2^{+12.0}_{-11.9}$ \\
\hline
100 & $3220^{+500}_{-580}$ & $144 .2^{+11.0}_{-11.0}$ \\
\hline
200 & $2690^{+430}_{-510}$ & $123 .6^{+9.6}_{-9.6}$ \\
\hline
500 & $2070^{+380}_{-470}$ & $98 .3^{+7.9}_{-7.9}$ \\
\hline
1000 & $1680^{+330}_{-440}$ & $81 .6^{+6.7}_{-6.7}$  \\
\hline
\end{tabular}
\caption{\label{table}Bounds on the suppression scale $M_\ast$ at $95\%\,{\rm CL}$ derived from the latest XENON100 data. The quoted errors reflect the scale uncertainties.}
\end{table}

From the cross sections~(\ref{eq:crosssection})  we can calculate the differential event rate in a direct detection experiment using
\beq
\label{eq:eventrate}
\frac{dR}{dE_\mathrm{R}} = \frac{\rho_0}{m_N m_\chi} \int_{v_\text{min}} \! d^3 v \; v f(v) \, \frac{d\sigma}{dE_\mathrm{R}}\left(v, E_\mathrm{R}\right) \, ,
\eeq 
where $\rho_0 = 0.3\, \text{GeV/cm}^3$ is the local DM density, $f(v)$ is the corresponding velocity distribution and $v_\text{min} = (m_\chi + m_N)/m_\chi \sqrt{E_\mathrm{R} / (2 \hspace{0.25mm} m_N)}$. We assume that the DM velocity distribution is a Maxwell-Boltzmann distribution with velocity dispersion $v_0 = 220 \, \text{km/s}$ and a cut-off at the escape velocity $v_\text{esc} = 544\, \text{km/s}$. Substituting~(\ref{eq:crosssection}) into~(\ref{eq:eventrate}) and calculating the velocity integral (see {\it e.g.} \cite{Barger:2010gv} for details), we can translate bounds on the differential event rate from direct detection experiments into bounds on the magnetic and electric dipole moments ${\cal C}_M/M_\ast^2$ and ${\cal C}_E/M_\ast^2$. Using~(\ref{eq:MLL}) for ${\cal C}_M$ and an analogue equation for ${\cal C}_E$, we can further convert these bounds into limits on the new-physics scale $M_\ast$ entering the definition of the operators $\mathcal{O}_T$ and $\mathcal{O}_{PT}$. In our analysis  we constrain the suppression scale using results from CRESST-I~\cite{Altmann:2001ax}, CDMS-II~\cite{Ahmed:2010wy}, PICASSO~\cite{Archambault:2012pm} and XENON100~\cite{Aprile:2012nq}.

\subsection{Hadron collider bounds}
\label{sec:colliderbounds}

\begin{figure}[!t]
\centering
\includegraphics[width=0.675 \textwidth, clip, trim = 7 0 -7 0]{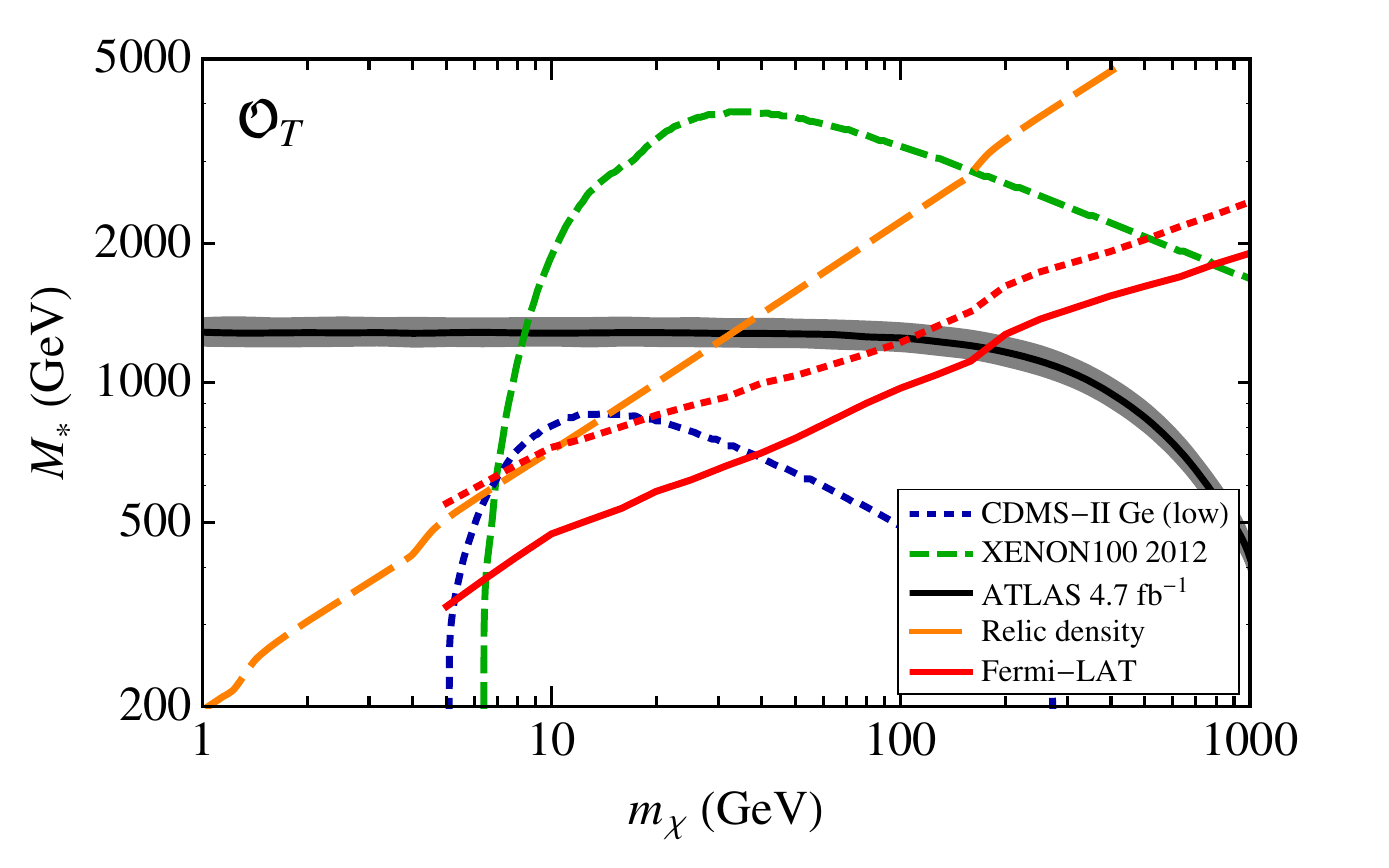}
\includegraphics[width=0.675 \textwidth]{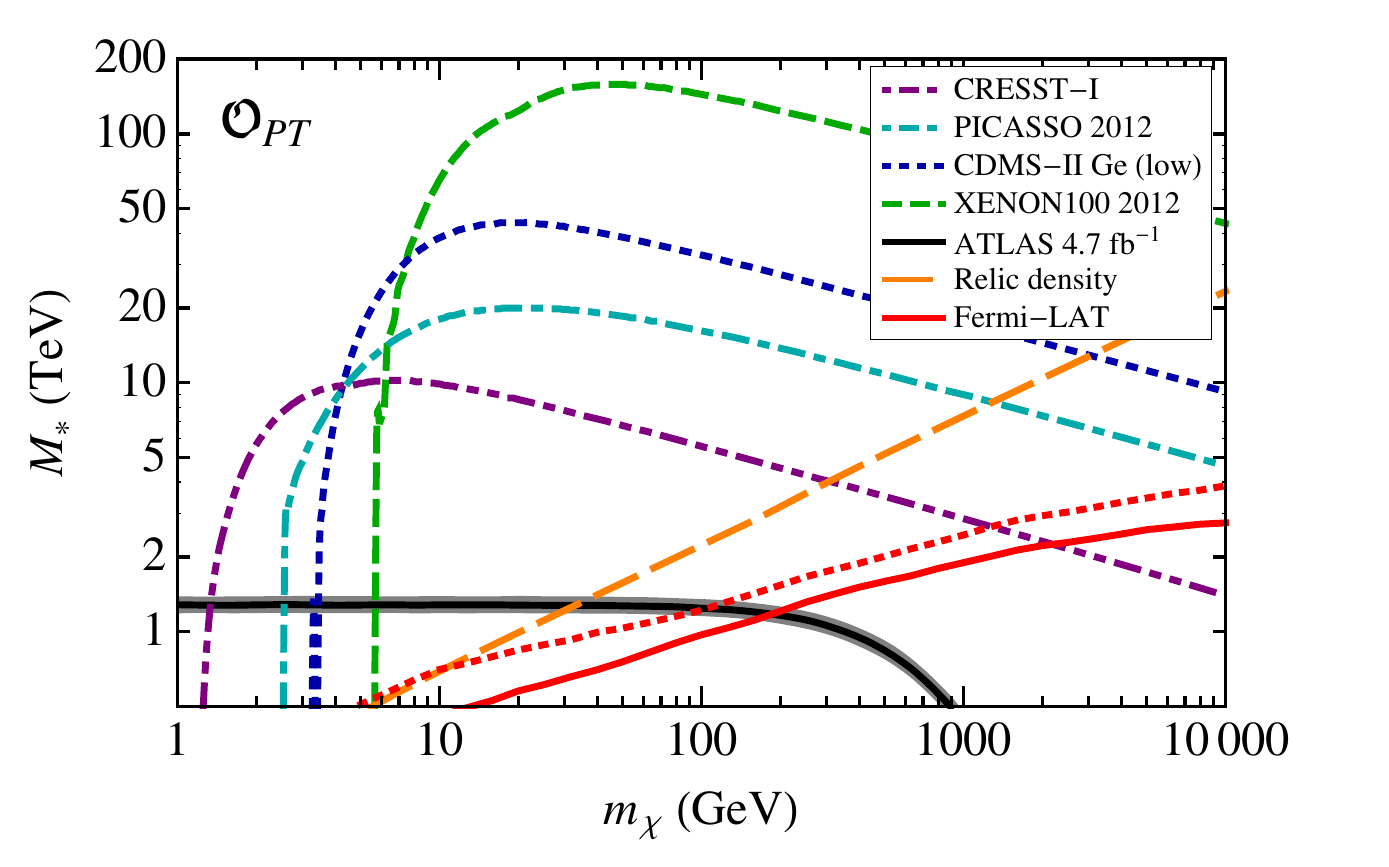}
\caption{\label{fig:Magneticbound} Upper panel (Lower panel): bounds on the scale $M_\ast$ suppressing the effective operator $\mathcal{O}_{T}$ ($\mathcal{O}_{PT}$) inferred from constraints on the DM magnetic dipole moment, LHC searches for jets plus MET,  and Fermi-LAT diffuse $\gamma$-ray measurements. The orange (long dashed) curve indicates the requirement for the correct relic density. For Fermi-LAT we show bounds with (dotted) and without (solid) modelling of the astrophysical foreground. The width of the collider limit indicates scale uncertainties, while the corresponding uncertainties for the dipole moments are not shown. For XENON100 the latter errors can be found in Tab.~\ref{table}.}
\end{figure}

In order to determine the cross section for a jet plus MET signal associated to the operators in~(\ref{eq:SD}) and~(\ref{eq:SDsup}), we have implemented each of them in {\tt FeynRules}~\cite{Christensen:2008py}. The actual computations have been performed at leading order with {\tt MadGraph~$\!\!5$}~\cite{Alwall:2011uj} utilising {\tt CTEQ6L1} parton distributions~\cite{Pumplin:2002vw}. We adopt the cuts from the most recent monojet search at ATLAS with an integrated luminosity of $4.7\,\text{fb}^{-1}$ at $\sqrt{s} = 7\,\text{TeV}$~\cite{ATLAS:2012ky}. For the signal region 4, comprising events with $\slashed{E}_T, p_T(j_1) > 500\,\text{GeV}$, the ATLAS results exclude new contributions to the production cross section in excess of $6.9\,\text{fb}$ at $95\% \,  {\rm CL}$.

In order to assess the theoretical errors in our analysis, we have studied the scale ambiguities of our results. Following common practice, we have set the renormalisation and factorisation scales equal to each other ($\mu = \mu_R = \mu_F $) and varied $\mu$ around $(\slashed E_T)_\text{min}=500 \, {\rm GeV}$ by a factor of 2. The resulting scale uncertainties amount to about $\pm30 \%$ at the cross-section level. We do not consider the effects of parton showering and hadronisation in our computations, but we do include secondary jets. The next-to-leading order (NLO) rate for the operator ${\cal O}_{AX}$ has recently been computed~\cite{Fox:2012ru} and results in a modest increase of around $40\%$ in the monojet cross section. NLO corrections of similar size are expected for ${\cal O}_{T}$, ${\cal O}_{PT}$ and ${\cal O}_{AN}$, but ignored here for simplicity, as they do not change our main conclusions qualitatively.

Our results are summarised in Tab.~\ref{table} and displayed in Fig.~\ref{fig:Magneticbound}. The constraints on the DM dipole moments from direct detection experiments allow to probe and exclude suppression scales $M_\ast$ in the range of around $[1.7, 3.7] \, {\rm TeV}$ for the tensor operator (upper panel) and in the range of $[66, 159] \, {\rm TeV} $ for the pseudotensor operator (lower panel). The black solid curves and grey bands in the two figures correspond to the $95\% \, {\rm CL}$ limits from the ATLAS $4.7 \, {\rm fb}^{-1}$~\cite{ATLAS:2012ky} search for a jet plus MET signal. For DM masses $m_\chi \gtrsim 10 \, {\rm GeV}$ the resulting bounds are clearly weaker than the constraints obtained from direct detection experiments.

\subsection{Relic density requirements}
\label{sec:relicbounds}

Further restrictions on the DM parameter space can be derived by calculating the thermal relic density~\cite{Kolb:1990vq}
\beq \label{eq:relic}
\Omega_{\rm DM} h^2 \simeq \frac{1.07 \cdot 10^9}{\rm GeV} \frac{2 \hspace{0.25mm} x_F}{M_{\rm Pl} \hspace{0.25mm} \sqrt{g_\ast}  \left (  a + \frac{3b}{x_F} \right )} \,.
\eeq 
In the above formula, $M_{\rm Pl} \simeq 1.22 \cdot 10^{19} \, {\rm GeV}$ denotes the Planck mass, $g_\ast \in [80,90]$ is the number of relativistic degrees of freedom at the freeze-out and $x_F = m_{\chi}/T_F \in [20, 30]$ with $T_F$ representing the freeze-out temperature. The factor of 2 in the numerator results from averaging for Dirac fermions.

The coefficients $a$ and $b$ entering (\ref{eq:relic}) are determined from the corresponding thermal-averaged annihilation cross section via $\langle \sigma v \rangle = a + 6b \hspace{0.5mm} T/m_\chi + {\cal O} ( T^2/m_\chi^2 )$. For the effective tensor ${\cal O}_T$ and pseudotensor ${\cal O}_{PT}$ interactions, we find the following $s$-wave coefficients 
\beq
a_{T} = \frac{6 m_\chi^2}{\pi M_\ast^4} \sum_f \sqrt{1 - z_f} \left (1 + 2 z_f \right ) \,, \qquad 
a_{PT} = \frac{6 m_\chi^2}{\pi M_\ast^4} \sum_f \left (1 - z_f \right )^{3/2}\,, 
\label{eq:annihilation}
\eeq
where the sum extends over all quark flavours $f$ with a mass $m_f < m_\chi$ and $z_f = m_f^2/m_{\chi}^2$. The above results agree with the expressions presented in~\cite{MarchRussell:2012hi}.

If the interactions due to ${\cal O}_T$ and ${\cal O}_{PT}$ are the only ones at freeze-out, one can determine the suppression scale $M_\ast$ for fixed $m_\chi$ by the requirement to obtain the observed DM relic abundance $\Omega_{\rm DM} h^2 \simeq 0.11$~\cite{Hinshaw:2012fq}. The values of $M_\ast$ which fulfil this constraint are indicated by the orange dashed curves in Fig.~\ref{fig:Magneticbound}. The parameter space above (below) this curves corresponds to DM overproduction (underproduction). 
Combining the direct detection with the relic density constraints  then enables one to derive  a lower bound on the DM mass. We find 
\beq \label{eq:absbounds} \
{\cal O}_T: \; \; m_\chi \gtrsim 160 \, {\rm GeV} \, , \qquad 
{\cal O}_{PT}: \; \; m_\chi \gtrsim 23 \, {\rm TeV} \, .
\eeq
Note that the parameter space that leads to DM underproduction in standard freeze-out can still give the observed relic abundance if the dark sector carries an initial asymmetry similar to the one in the visible sector~\cite{MarchRussell:2012hi, Kaplan:1991ah, Kaplan:2009ag}.  Since the presence of any asymmetric component necessarily leads to a larger relic density compared to the purely symmetric case,  the bounds~(\ref{eq:absbounds}) are conservative lower limits even in theories of asymmetric DM.

\subsection{Bounds from indirect detection}
\label{sec:indirectbounds}

Another bound on the new-physics scale $M_\ast$ can be derived from the Fermi-LAT measurements of diffuse $\gamma$-ray emission~\cite{Ackermann:2012rg}. Since $\gamma$-rays are unavoidable products of hadronisation, these measurements constrain the DM annihilation cross section into quarks. Using~(\ref{eq:annihilation}), we can directly translate the results from~\cite{Ackermann:2012rg} into bounds on $M_\ast$ after applying a rescaling factor of $(0.43 / 0.3)^2 \simeq 2$ to account for our choice of local DM density (see also~\cite{Cirelli:2009dv, Cheung:2011nt}). The resulting limits are shown as red curves in Fig.~\ref{fig:Magneticbound}. The solid curves correspond to the conservative limit without background modelling, while the dotted lines indicate the limits derived in~\cite{Ackermann:2012rg} by modelling the foreground astrophysical diffuse emission. For the tensor operator and DM masses above approximately $[500, 800] \, {\rm GeV}$, these limits are stronger than those obtained both from direct detection and collider experiments. In this region, however, significantly larger values of $M_\ast$ are required to obtain the correct DM density.

Effective operators that give rise to tree-level annihilation of DM into quarks typically induce annihilation into mono-energetic $\gamma$-rays at the loop level, potentially providing an explanation for the claimed observation of a signal at around $130 \, {\rm GeV}$~\cite{Bringmann:2012vr, Weniger:2012tx}. However, in order to obtain a $\gamma$-ray flux close to the current sensitivity of Fermi-LAT~\cite{Ackermann:2012qk}, the suppression scale $M_\ast$ has to be around $400 \, {\rm GeV}$ or smaller~\cite{Goodman:2010qn}, which is well below our bounds. In fact, as was pointed out in~\cite{Buchmuller:2012rc, Cohen:2012me, Cholis:2012fb}, such a scenario is already excluded just from limits on the diffuse $\gamma$-ray emission. Furthermore, it turns out that for the specific case of tensor and pseudotensor operators, the one-loop contribution for annihilation into $\gamma$-rays vanishes identically~\cite{Goodman:2010qn} (see also \cite{Bobeth:2011st} for a discussion in a non-DM context). The leading contributions to $\bar \chi  \chi \to \gamma \gamma$ then arise at two loops from one-particle reducible diagrams involving two insertions of  ${\cal O}_M$ or  ${\cal O}_E$.  As a result the associated  fluxes of mono-energetic $\gamma$-rays  are suppressed by two powers of (\ref{eq:MLL}), rendering them undetectably small.

\subsection{Direct detection bounds on SD interactions}
\label{sec:SDbounds}

\begin{figure}[!t]
\centering
\includegraphics[width=0.675 \textwidth]{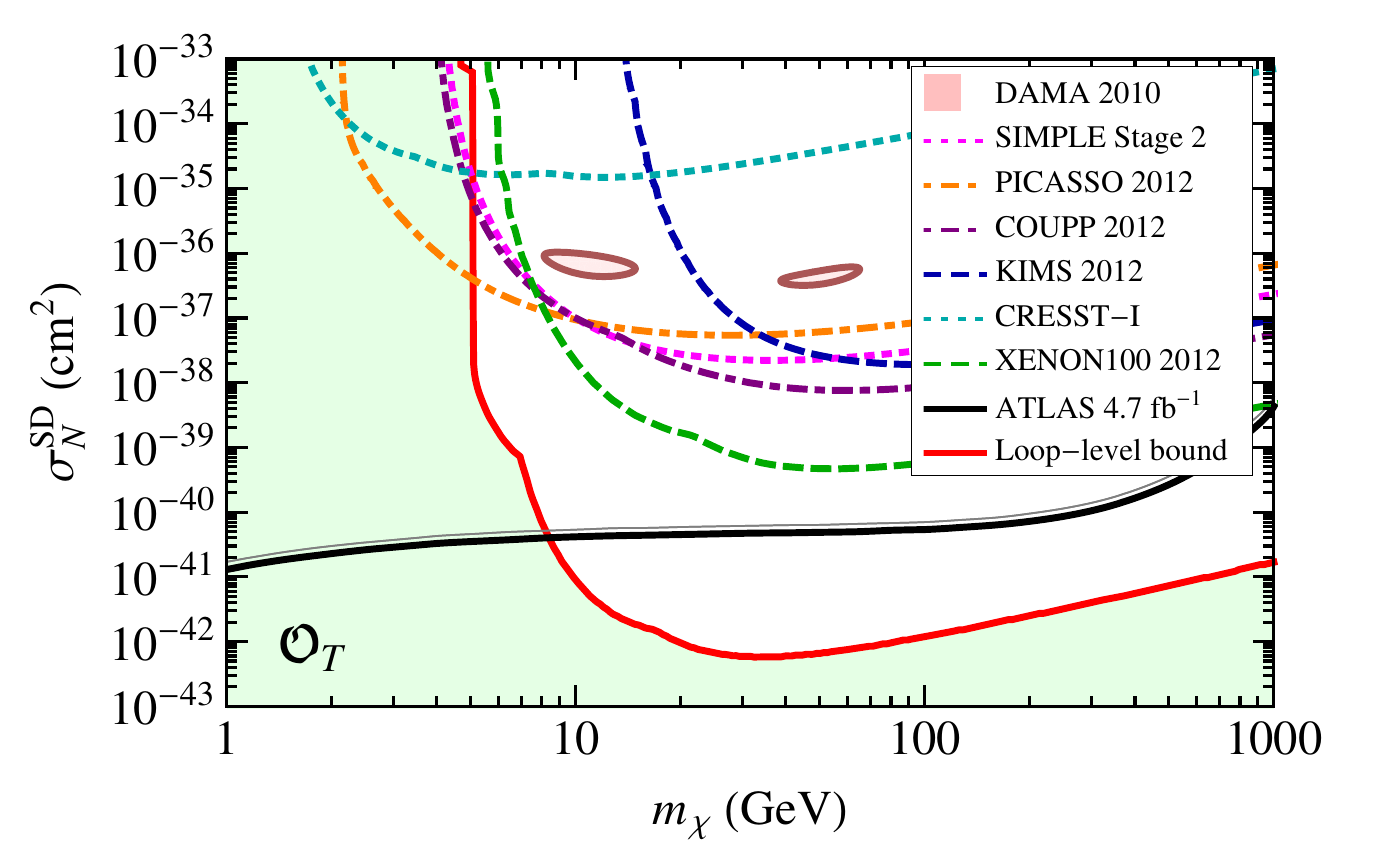}
\caption{\label{fig:MagneticDD} Bounds on the SD DM-nucleon cross section from direct detection experiments compared to the combined constraint on the operator ${\cal O}_T$ resulting from loop-induced DM dipole moments and LHC searches for jets plus MET. The width of the collider limit indicates scale uncertainties, while the corresponding uncertainties for the dipole moments are not shown.}
\end{figure}

For the tensor operator  we can translate the constraints on ${\cal O}_T$ derived in Secs.~\ref{sec:dipolebounds} to~\ref{sec:indirectbounds} into a limit on the SD DM-nucleon cross section using the formula
\begin{equation}
\sigma^\text{SD}_N = a_N^2 \frac{12}{\pi}\frac{m_\text{red}^2}{M_\ast^4} \,,
\label{eq:SDDD}
\end{equation}
where $m_\text{red} = m_\chi m_N/(m_\chi + m_N)$ is the reduced mass of the DM-nucleon system and
\begin{equation}
a_N = \sum_{q=u,d,s} \Delta q^{(N)} \,,
\end{equation}
is the effective coupling of the DM particle to the nucleon spin assuming flavour independent quark couplings. The coefficients $\Delta q^{(N)}$ encode the contributions of the light quark to the nucleon spin and can be extracted from polarised deep inelastic scattering. Using the PDG values $\Delta u^{(p)} = \Delta d^{(n)} = 0.84 \pm 0.02$, $\Delta d^{(p)} = \Delta u^{(n)} = -0.43 \pm 0.02$ and $\Delta s^{(p)} = \Delta s^{(n)} = -0.09 \pm 0.02$ \cite{Beringer:1900zz}, we find, ignoring errors, $a_p \simeq a_n \simeq 0.32$.

We can now compare these indirect bounds on $\sigma_N^\text{SD}$ with the constraints on SD interactions from direct detection experiments with an odd number of protons or neutrons in the target atoms. Such direct constraints stem from CRESST-I~\cite{Altmann:2001ax}, SIMPLE~\cite{Felizardo:2011uw}, PICASSO~\cite{Archambault:2012pm}, COUPP~\cite{Behnke:2012ys}, KIMS~\cite{KIMS} and XENON100~\cite{Aprile:2012nq}. For XENON100, we calculate bounds using the ``maximum gap" method~\cite{Yellin}, for all other experiments we use the ``binned Poisson" method~\cite{Green:2001xy, Savage:2008er}. The DAMA region is calculated using a $\chi^2$ parameter estimation method under the assumption of no channelling. We take the SD form factors from~\cite{Menendez:2012tm} for xenon, from~\cite{Toivanen} for caesium and from~\cite{Bednyakov:2006ux} for all other target materials.

The resulting bounds and best-fit regions are shown in Fig.~\ref{fig:MagneticDD}. The solid red curve indicates the combined bound from loop-induced SI interactions. It is obvious from these results that if DM interactions proceed via the operator $\mathcal{O}_T$, experiments sensitive for SD interactions cannot currently probe the allowed parameter region. In fact, even if we neglect the bounds from the LHC, the accessible parameter region is already excluded by constraints on the DM magnetic dipole moment unless $m_\chi < 5 \, {\rm GeV}$. We do not show the corresponding plot for the pseudotensor operator, because in this case tree-level scattering is momentum suppressed and therefore the corresponding bounds from direct detection experiments do not give any interesting constraint. All relevant bounds for $\mathcal{O}_{PT}$ are already shown in Fig.~\ref{fig:Magneticbound}. We note that the effective interactions $\mathcal{O}_T$ and $\mathcal{O}_{PT}$ cannot provide an explanation of the DAMA/LIBRA modulation~\cite{Bernabei:2010mq}. 

To conclude this section,  let us consider the contribution of bottom and charm quarks to DM dipole moments separately. These contributions are suppressed compared to the one from top quarks because of the smaller quark masses, but they will receive larger logarithmic corrections. For the tensor operator ${\cal O}_T$, we find numerically that the resulting bounds on $M_\ast$ are weaker compared to the ones given in Tab.~\ref{table} by a factor of approximately $0.14$ for bottom quarks and a factor of approximately $0.12$ for charm quarks. For the pseudotensor operator ${\cal O}_{PT}$, the limits  are weaker by a factor of $0.12$ and $0.10$, respectively. Consequently, in the latter case both bottom- and charm-quark loops alone lead to bounds that significantly exceed the constraints from both the LHC and direct detection.

\section{Constraints on axialvector and anapole operators}
\label{sec:axial}
\begin{figure}[!t]
\begin{center}
\includegraphics[width=0.67\columnwidth,clip,trim=0 0 0 0]{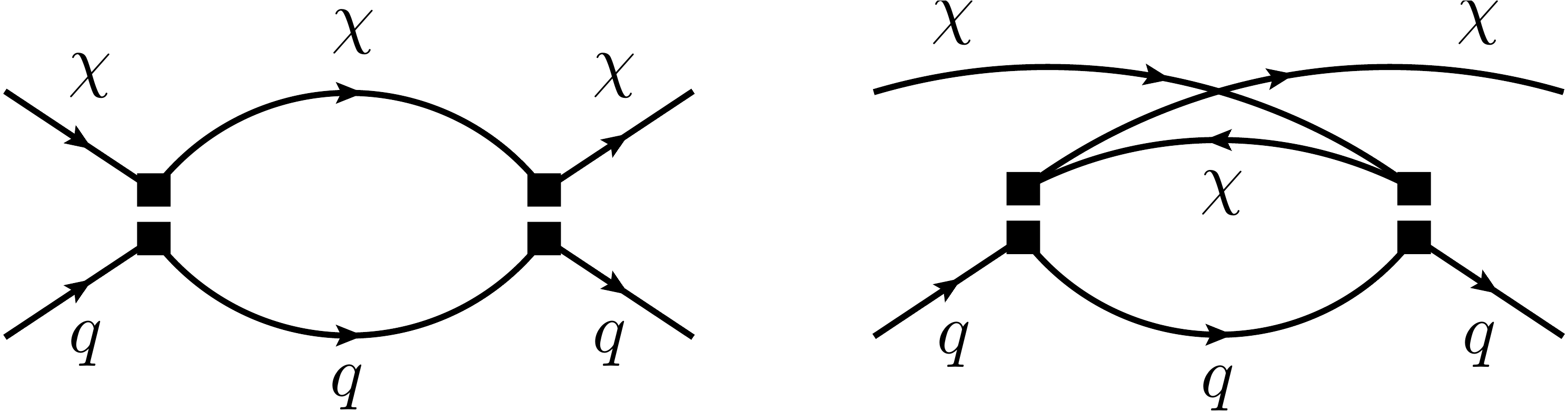}
\vspace{1mm}
\caption{\label{fig:diagramsloop} One-loop contributions to the DM-quark scattering amplitude induced by the operators $\mathcal{O}_{AX}$ and $\mathcal{O}_{AN}$. The black squares represent operator insertions. }
\end{center}
\end{figure}

We now turn to the discussion of the axialvector operator $\mathcal{O}_{AX}$ and the anapole operator $\mathcal{O}_{AN}$. In both cases, we consider the one-loop diagrams shown in Fig.~\ref{fig:diagramsloop}.  As has been observed previously in the literature~\cite{Cirelli:2005uq,Essig:2007az,Kopp:2009et,Freytsis:2010ne}, the matrix element resulting from two insertions of the operator $\mathcal{O}_{AX}$ leads to SI scattering, which can be described by the operator $\mathcal{O}_S$ defined in~(\ref{eq:SI}).\footnote{The Feynman diagrams in  Fig.~\ref{fig:diagramsloop} also lead to a loop-suppressed contribution to $\mathcal{O}_{AX}$, but a correction to the vector operator $\mathcal{O}_{V}$ is not introduced.} Assuming for definiteness $m_\chi < m_t$, the induced coefficient reads  (see App.~\ref{app:1} for details)
\begin{equation}
\label{eq:axialloop}
{\cal C}_S \simeq -\frac{1}{2 \pi^2 } \hspace{0.25mm} \frac{m_\chi}{M_\ast} \ln  \frac{M_\ast^2}{m_\chi^2} \, ,
\end{equation}
for $q = u,d,s,c,b$ while for $q= t$ the argument of the logarithm is $M_\ast^2/m_t^2$. The same leading-logarithmic result (though with opposite overall sign) holds if one considers two insertions of $\mathcal{O}_{AN}$ instead of $\mathcal{O}_{AX}$. As in the case of (\ref{eq:MLL}) the large logarithm appearing in~(\ref{eq:axialloop}) can be resummed using RG methods. The effect of such a resummation turns out to have a minor numerical effect of $10\%$ or below.  Since they do not change our final results qualitatively, we will ignore RG effects in what follows.

The SI scattering cross section for the scalar operator takes the form 
\begin{equation}
\label{eq:SIDD} 
\sigma^\text{SI}_N =  \frac{f_N^2}{\pi} \hspace{0.25mm} \frac{m_\text{red}^2 m_N^2}{M_\ast^6} \, {\cal C}_S^2 \,.
\end{equation}
The coefficient $f_N$ is the effective DM-nucleon coupling, which is given by
\begin{equation} \label{eq:fN} 
f_N \simeq \sum_{q=u,d,s} f_{T_q}^{(N)} + \frac{4}{27} \,   f_{T_G}^{(N)} \,,
\end{equation}
assuming $M_\ast < m_t$. The light-quark matrix elements $f_{T_q}^{(N)}$ can either be determined phenomenologically from baryon masses and meson-baryon scattering data or computed within lattice QCD. We adopt the values $f_{T_u}^{(N)} \simeq 0.021$ and $f_{T_d}^{(N)} \simeq 0.038$ from~\cite{Freytsis:2010ne} for up  and down quarks and  $f_{T_s}^{(N)} \simeq 0.013$ from~\cite{Takeda:2010id} for strange  quarks. The gluon matrix element is defined as $f_{T_G}^{(N)} = 1 - \sum_{q=u,d,s} f_{T_q}^{(N)}$.
For $M_\ast > m_t$, there is an additional contribution from the top quark
\begin{equation} 
\Delta f_N = \frac{2}{27} \,  f_{T_G}^{(N)} \, \frac{ \ln \left( M_\ast^2 / m_t^2 \right) }{ \ln \left( M_\ast^2 / m_\chi^2 \right) } \, ,
\end{equation}
which turns out to be numerically subleading. As a result the SI scattering cross section~(\ref{eq:SIDD}) becomes essentially independent of the top-quark contribution and we find $f_p \simeq f_n \simeq 0.21$.

\subsection{Bounds on the new-physics scale}

\begin{figure*}[!t]
\centering
\includegraphics[width=0.675 \textwidth]{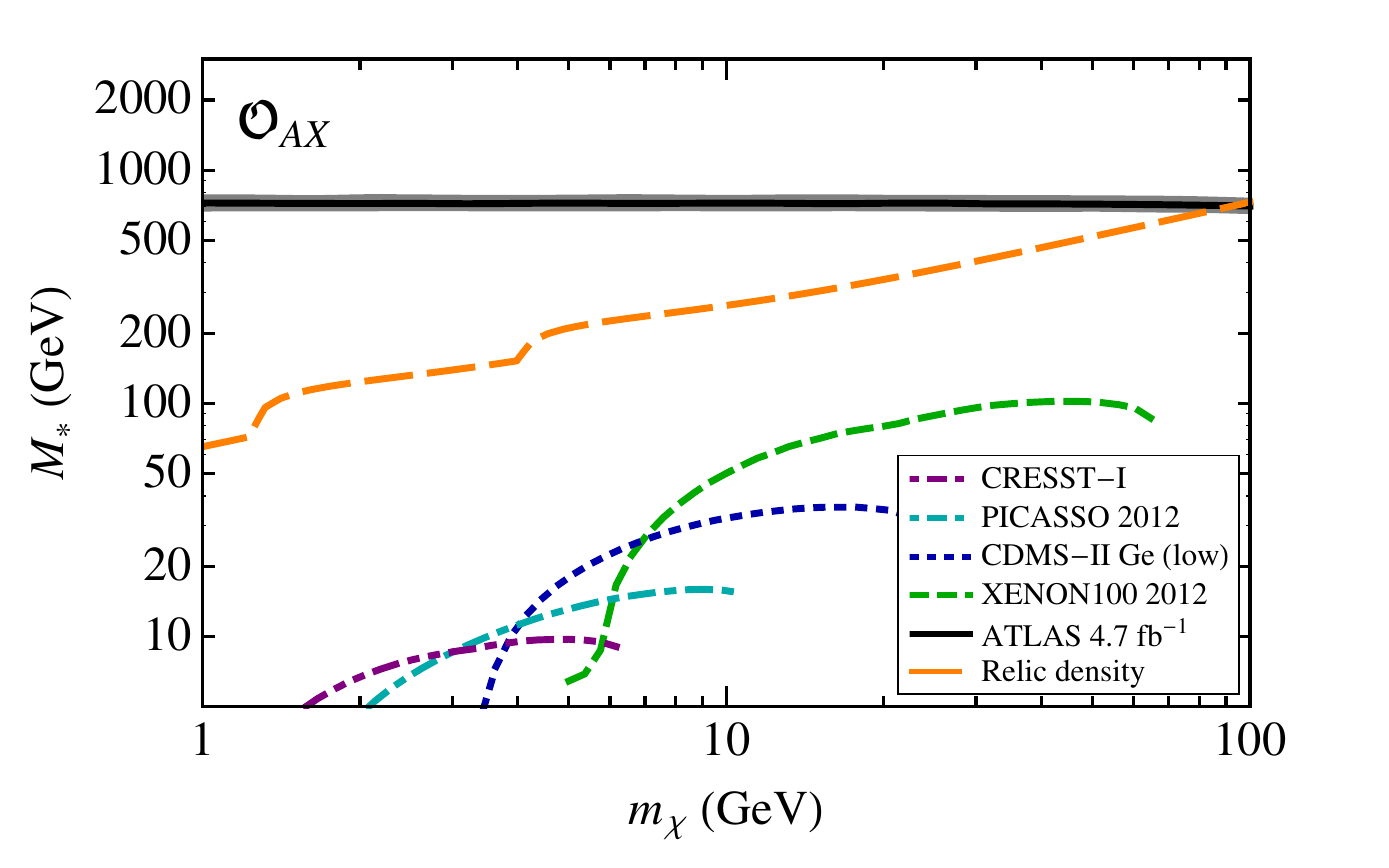}
\includegraphics[width=0.675 \textwidth]{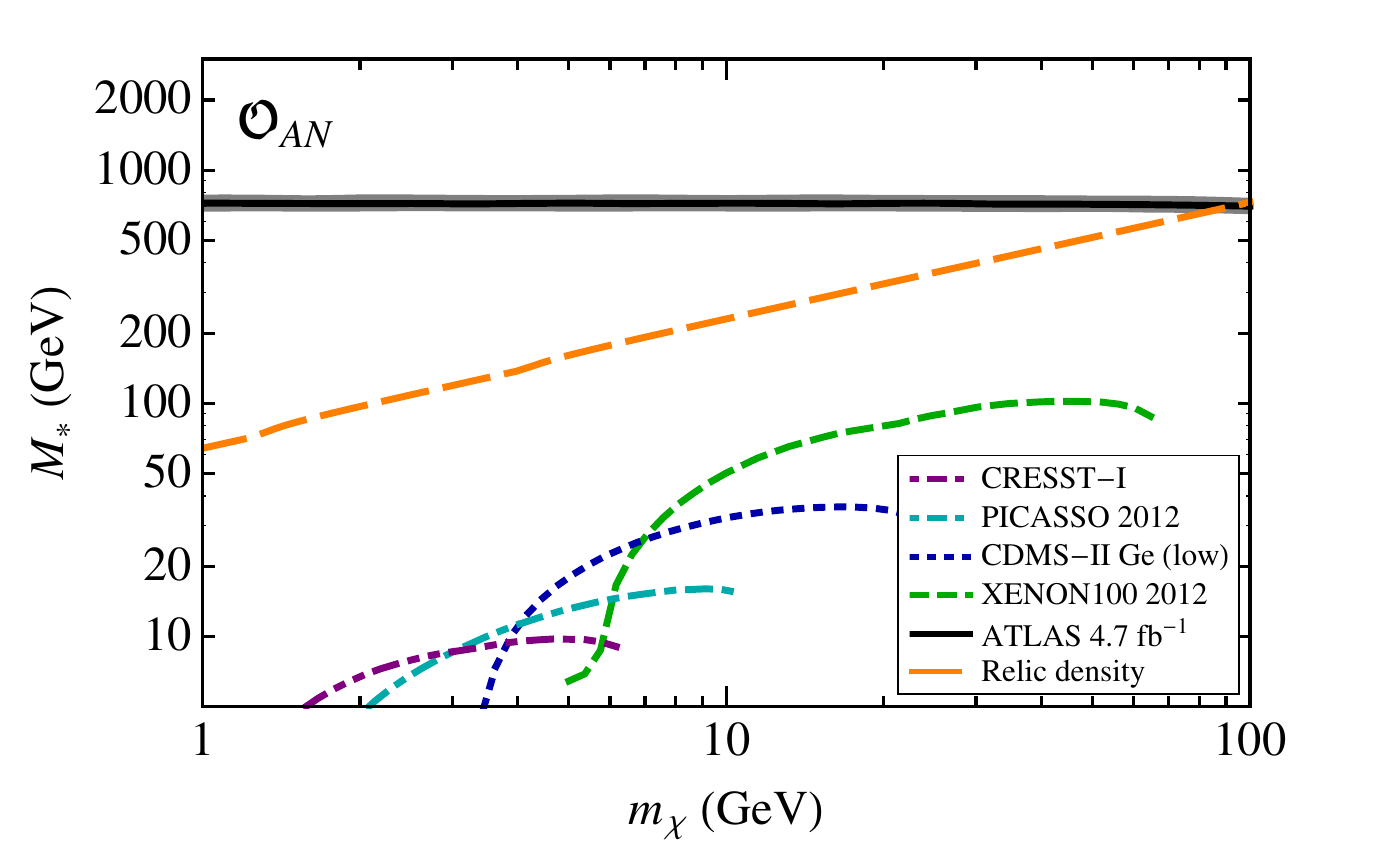}
\caption{\label{fig:Axialbound} Bounds on the scale $M_\ast$ suppressing the effective operators $\mathcal{O}_{AX}$ (top) and $\mathcal{O}_{AN}$~(bottom) inferred from LHC monojet searches and constraints on the SI DM-nucleon cross section. The orange~(long dashed) curve in each plot indicates the requirement for the correct relic density. The width of the collider limit indicates scale uncertainties.}
\end{figure*}

The scattering cross section for SI interactions is strongly constrained by direct detection experiments (see {\it e.g.}~\cite{Ahmed:2009zw,Ahmed:2010wy,Felizardo:2011uw,Archambault:2012pm,Aprile:2012nq}). Inserting the bounds on $\sigma^\text{SI}_N$ into (\ref{eq:SIDD}) one can constrain ${\cal C}_S$ and therefore, using (\ref{eq:axialloop}), the scale $M_\ast$ appearing in $\mathcal{O}_{AX}$ and $\mathcal{O}_{AN}$. Our results are shown in Fig.~\ref{fig:Axialbound}.  For DM masses between $10 \, {\rm GeV}$ and $100 \, {\rm GeV}$, we find bounds in the range of $[60, 125] \, {\rm GeV}$. For smaller values of $m_\chi$, the bounds are suppressed due to the multiplicative factor $m_\chi$ in (\ref{eq:axialloop}), while for larger values the limits worsen because $m_\chi$ approaches $M_\ast$ and consequently the logarithm in (\ref{eq:axialloop}) tends to zero. In practice, we only show bounds that satisfy $M_\ast^2 / m_\chi^2 > 2$. Comparing these results to the constraints from LHC searches, we find that the latter give far superior bounds, requiring $M_\ast \gtrsim 700 \, {\rm  GeV}$.

As before we also calculate the thermal-averaged annihilation cross section into quarks. For the axialvector operator ${\cal O}_{AX}$ we find  that $s$-wave annihilation into pairs of quarks is helicity suppressed, so that one has to include the contribution from $p$-wave annihilation. Explicitly, we get 
\beq \label{eq:abav} 
a_{AX} = \frac{3}{2 \pi M_\ast^4} \sum_f m_f^2 \hspace{0.25mm} \sqrt{1 - z_f} \,, \qquad 
b_{AX} = \frac{3 m_\chi^2}{2 \pi M_\ast^4} \sum_f  \hspace{0.25mm} \sqrt{1 - z_f} \; \frac{8  - 22 z_f + 17 z_f^2 }{24 \left (1 - z_f \right)} \,. 
\eeq
In the case of the anapole operator ${\cal O}_{AN}$ we observe  that annihilation can only proceed via $p$-wave. The corresponding coefficients are given by
\beq \label{eq:aba}
a_{AN} = 0 \,, \qquad b_{AN} = \frac{m_\chi^2}{4 \pi M_\ast^4} \sum_f \sqrt{1 - z_f} \left  (2 + z_f \right )\,.
\eeq
Our results (\ref{eq:abav}) and (\ref{eq:aba})  agree with those given in \cite{MarchRussell:2012hi}. 
Since annihilation is suppressed for both operators, $M_\ast$ has to be significantly smaller than for the tensor and pseudotensor operator in order to reproduce the observed relic density. LHC bounds then imply that the required DM density can only be achieved for  $m_\chi \gtrsim 93 \, {\rm GeV}$ for both the axialvector and the anapole operator.

Because of the helicity suppression for the axialvector operator, constraints from Fermi-LAT are very weak for $m_\chi < m_t$. To calculate these constraints accurately, one would have to take into account both loop-induced annihilation into monoenergetic $\gamma$-rays as well as photons arising from the process $\bar{\chi} \chi \rightarrow \bar{q} q V$, where $V$ is a SM gauge boson (see~{\it e.g.}~\cite{Bell:2011if}). We do not perform such a calculation here, because the resulting bounds turn out to be  much weaker than existing collider constraints~\cite{Goodman:2010qn,Cheung:2011nt}.  For the anapole operator, the processes $\bar{\chi} \chi \rightarrow \bar{q} q$ and $\bar{\chi} \chi \rightarrow \bar{q} q V$ are both velocity suppressed and the one-loop contribution to annihilation into $\gamma$-rays vanishes.\footnote{The latter finding is in line with the explicit calculation performed in \cite{Bobeth:2011st} but disagrees with the results presented in \cite{Goodman:2010qn}, where a bound on the anapole operator has been inferred using Fermi-LAT data.} Consequently, no relevant constraints on the operator ${\cal O}_{AN}$ arise from Fermi-LAT.

\subsection{Bounds on the DM scattering cross section}

\begin{figure*}[!t]
\centering
\includegraphics[width=0.675 \textwidth]{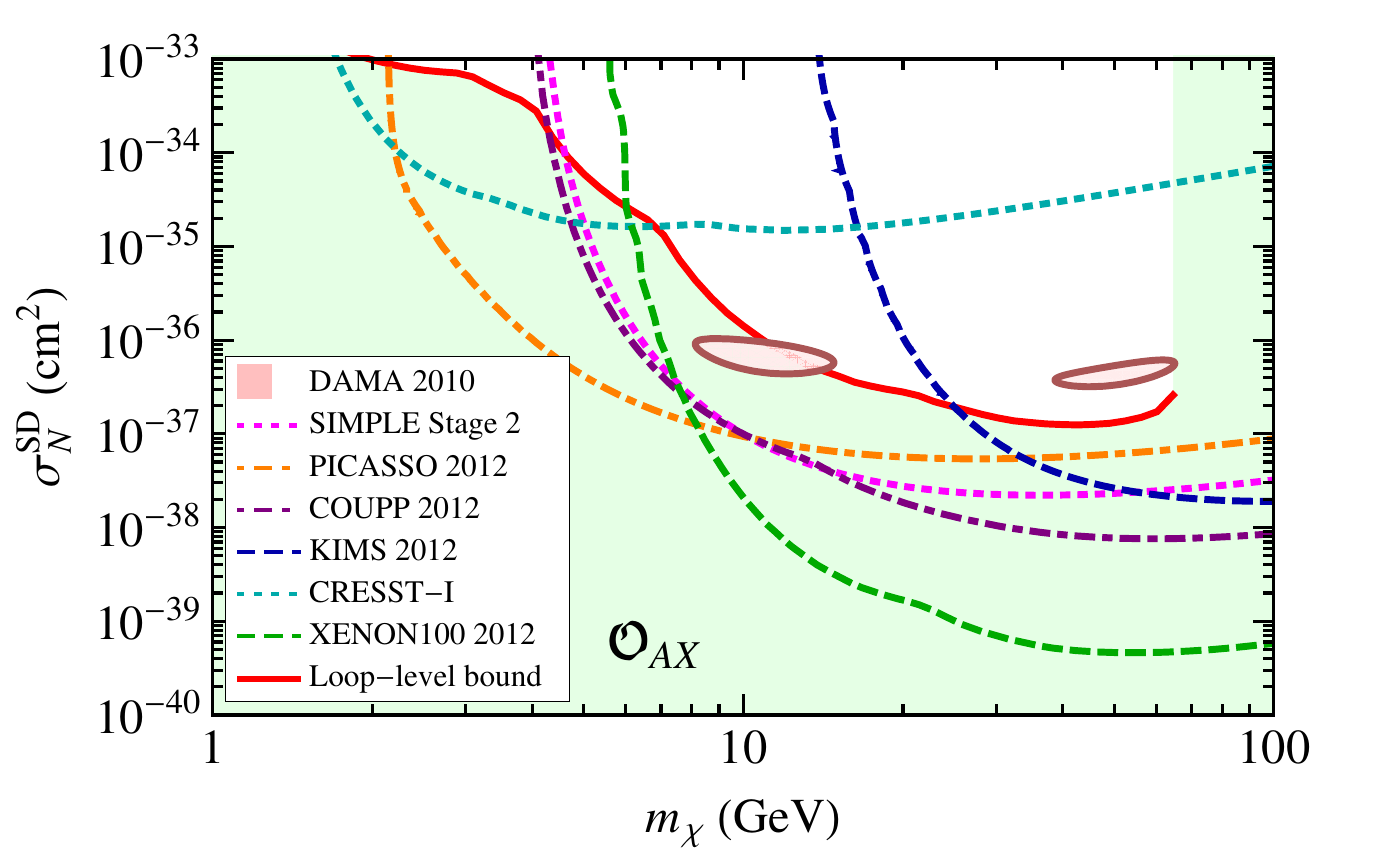}
\includegraphics[width=0.675 \textwidth]{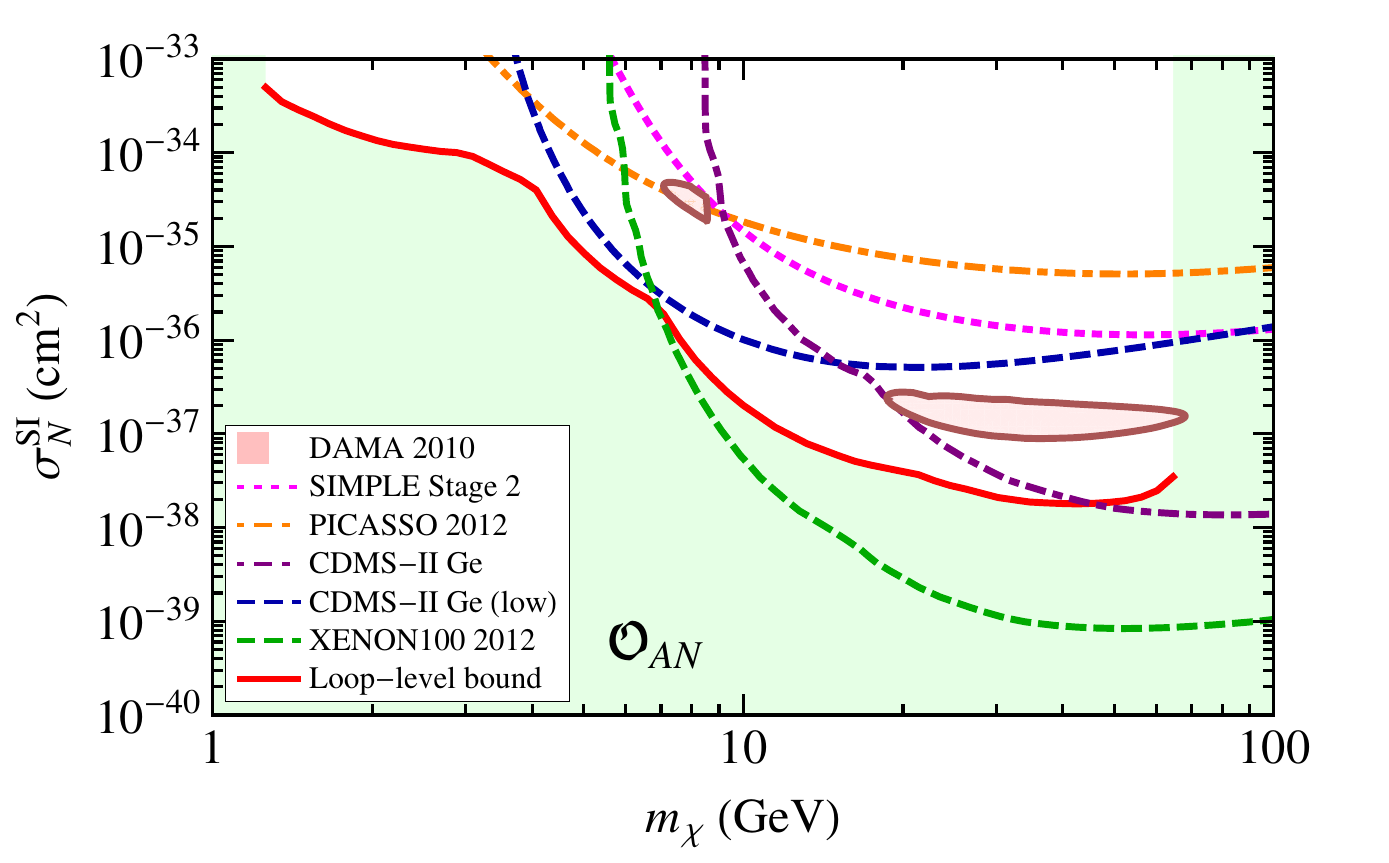}
\caption{\label{fig:Anapolebound} Upper panel: bounds on the SD scattering cross section from tree-level processes compared to the bound from loop-induced SI interactions (solid red curve) for the axialvector operator $\mathcal{O}_{AX}$. Lower panel: corresponding limits on the SI reference cross section for the anapole operator $\mathcal{O}_{AN}$.}
\end{figure*}

Even though Fig.~\ref{fig:Axialbound} seems to indicate that loop-induced SI scattering cross sections do not lead to relevant constraints on the scale $M_\ast$, it is interesting to consider cases where the bounds from LHC searches and the relic density requirement do not apply, for example low-mass asymmetric DM interacting with the SM via a mediator that is light compared to LHC scales~\cite{MarchRussell:2012hi}. In this case, only direct detection experiments can give lower limits  on $M_\ast$. In this section, we will therefore directly compare the limits from loop-induced SI interactions with the bounds from tree-level SD interactions (for the axialvector operator  $\mathcal{O}_{AX}$) and momentum suppressed interactions (for the anapole operator $\mathcal{O}_{AN}$) to determine which contribution gives the most stringent constraint.   

For the axialvector operator $\mathcal{O}_{AX}$, the constraints on $M_\ast$ derived above can be translated into bounds on the SD scattering cross section using
\begin{equation}  
\sigma^\text{SD}_N = a_N^2 \hspace{0.25mm} \frac{3}{\pi}\frac{m_\text{red}^2}{M_\ast^4} \,,
\end{equation}
which differs from the corresponding formula for the tensor operator by a factor of $1/4$~\cite{Agrawal:2010fh}. Our results are shown in the upper panel of Fig.~\ref{fig:Anapolebound}. As before, the bounds are cut-off at large DM masses, because we require $M_\ast^2 / m_\chi^2 > 2$. Because of the small Yukawa couplings appearing in (\ref{eq:axialloop}) these constraints are relatively weak compared to the constraints from SD direct detection experiments. Nevertheless, for some targets with limited sensitivity to SD interactions, such as oxygen, silicon or germanium, the event rate resulting from loop-induced processes becomes comparable to tree-level scattering. Consequently, these effects should be included in a complete analysis of SD interactions.

Let us now turn to the anapole operator ${\cal O}_{AN}$. The differential scattering cross section is given by~\cite{Ho:2012bg}
\begin{equation}
\frac{d\sigma^\text{SI}}{d E_\mathrm{R}} = \frac{1}{2\pi} \frac{m_N }{M_\ast^4} \frac{Z^2}{v^2} \left[v^2 + \frac{q^2}{2 m_N^2} \left(1 - \frac{m_N^2}{2 m_\text{red}^2}\right)\right] \; .
\end{equation}
While this cross section is suppressed by powers of $v^2$ and $q^2 / m_N^2$, there is an enhancement proportional to $Z^2$ from coherence. Consequently, it is possible to directly constrain $M_\ast$ for the anapole operator using tree-level scattering in direct detection experiments. 

Again, the resulting constraints can be compared to the indirect bounds from loop-induced SI interactions. The resulting bounds  (in terms of $\sigma_N^\text{SI}$ as defined in (\ref{eq:SIDD})) are shown in the lower panel of Fig.~\ref{fig:Anapolebound}. We observe that, in contrast to the axialvector operator ${\cal O}_{AX}$, the loop-induced limits  for the case of~$\mathcal{O}_{AN}$ are comparable to the bounds obtained from tree-level interactions. In fact, the loop contributions are dominant for small DM masses because tree-level interactions are suppressed for small momentum transfer. In other words, in cases where the LHC bounds do not apply, loop-induced processes give the strongest bounds on the DM interactions in this specific region of parameter space.

\section{Conclusions}

With the first years of data taking at the LHC   searches for DM have made impressive progress and significantly improved their sensitivity. By now  both ATLAS and CMS provide stringent bounds on the cross section for jet plus MET signals. For effective operators that lead to either SD or momentum suppressed  scattering in low-energy experiments the sensitivity of the LHC seems to clearly exceed the reach of direct detection. In this paper, we have demonstrated that this is not necessarily true, because operators that give SD or  momentum suppressed interactions at tree level may induce SI interactions at loop level, which can be  efficiently constrained by direct detection experiments. These contributions have to be taken into account for a complete study of the sensitivity of DM direct detection.

In the present work, we have focused on effective four-fermion interactions. We first studied the tensor and pseudotensor operator. For these operators, heavy-quark loops induce magnetic and electric dipole moments, respectively, which are strongly constrained by direct detection. For DM masses $m_\chi \in [10 \, {\rm GeV}, 1 \, {\rm TeV}]$, these constraints allow to exclude new-physics scales $M_\ast$ in the range of around $[1.7, 3.7] \, {\rm TeV}$ ($[66, 159] \, {\rm TeV} $)  for the tensor (pseudotensor) operator.

In the case of the pseudotensor operator, the derived limits exceed by far all other experimental constraints on $M_\ast$. As a consequence, the contribution of this operator to DM annihilation into quarks has to be very small, so that DM would be overproduced in the early universe, unless either the DM mass is larger than $23 \, {\rm TeV}$ or additional annihilation channels are available. This result is an impressive demonstration both of the power of direct detection experiments as well as of the importance of loop contributions in calculating the expected scattering cross sections.

For the tensor operator, on the other hand, it is very interesting to compare the limits on $M_\ast$ from induced dipole moments to other ways of constraining the suppression scale. Collider searches for jets plus MET give $M_\ast \gtrsim 1.3 \, {\rm TeV}$ at small DM masses, providing the dominant constraints for $m_\chi \lesssim 10 \, {\rm GeV}$. In contrast, Fermi-LAT is most constraining for large DM masses, which yields the most stringent bound of  $M_\ast \gtrsim 2 \, {\rm TeV}$ for $m_\chi \gtrsim 500 \, {\rm GeV}$. Direct detection experiments that predominantly constrain SD interactions are not sensitive to any of the allowed parameter regions. In fact, the bounds on DM magnetic dipole moments from SI direct detection alone are sufficient to exclude almost the whole parameter space probed by SD searches. Finally, we point out that the combined constraints from collider searches, Fermi-LAT and direct detection exclude the possibility that the correct relic density is obtained for the tensor operator from thermal freeze-out for $m_\chi  \lesssim 160 \, {\rm  GeV}$ unless additional annihilation channels are present.

We then studied the axialvector and anapole operator. For these operators, SI interactions arise at the one-loop level from diagrams with two operator insertions. However, the induced effective quark couplings are proportional to the quark masses $m_q$, so the interactions of DM with light quarks are suppressed. Consequently, SI searches only give a relatively weak constraint on the suppression scale $M_\ast$. For the axialvector operator, these constraints are always weaker than constraints from SD searches. For the anapole operator scattering cross sections are typically smaller by an order of magnitude than the ones for the corresponding axialvector operator for the same values of $M_\ast$. Loop-induced interactions are hence more relevant in this case and provide the dominant contribution to the total event rate for low mass DM. 

Even stronger bounds for the axialvector and the anapole operator arise from collider searches, which typically require $M_\ast \gtrsim 700 \, {\rm GeV}$, and are therefore in conflict with the observed DM relic abundance unless $m_\chi \gtrsim 93 \,  {\rm GeV}$. Nevertheless, these bounds are significantly weaker than the corresponding ones for the tensor operator. As a consequence, even if scattering in direct detection experiments is SD, the sensitivity of future direct detection experiments can compete with indirect and collider searches.

Our study clearly demonstrates the remarkable complementarity of the different search strategies for DM. For many of the operators we consider, we find that direct detection, indirect detection and collider searches provide comparable bounds, and that different searches dominate for different values of the DM mass. Even in cases where one strategy seems to be inferior to the others when considering interactions only at tree level, loop contributions can change this picture drastically. Consequently, detailed studies of such loop effects will play an essential part in combining the virtues of all different search strategies in order to solve the DM problem.

\section*{Acknowledgements}
We are grateful to David Berge, Philipp Mertsch, Kai Schmidt-Hoberg, Steven Schramm, Thomas Schwetz-Mangold and Steven Worm for useful discussions and would like to thank Christopher McCabe for valuable comments on the manuscript. UH acknowledges travel support from the UNILHC network (PITN-GA-2009-237920). FK is supported by the Studienstiftung des Deutschen Volkes, the STFC and a Leathersellers' Company Scholarship. 

\appendix

\section{Calculation of one-loop amplitudes}
\label{app:1}

In this appendix we present explicit calculations for the one-loop diagrams in Figs.~\ref{fig:diagram}~and~\ref{fig:diagramsloop}. For an insertion of ${\cal O}_T$, the graph with a closed top-quark loop shown on the left in Fig.~\ref{fig:diagram}  yields the amplitude 
\beq \label{eq:A1}
{\cal A} = -\frac{i}{(2 \pi)^4} \frac{N_c  \hspace{0.25mm}  e \hspace{0.25mm} Q_t}{M_\ast^2} \left (  \bar \chi \hspace{0.25mm}  \sigma_{\mu \nu} \hspace{0.25mm}  \chi \right)\,  \mu^{4-d} \int \! d^d l \; \frac{{\rm tr} \left [   \sigma^{\mu \nu} \left ( \slashed{l} - \slashed {q} + m_t \right ) \slashed{\varepsilon} \left ( \slashed{l} + m_t \right )  \right ]}{\big ( \left ( l - q \right )^2 - m_t^2 \big ) \left ( l ^2 - m_t^2 \right ) } \,, 
\eeq
where we have used dimensional regularisation. The factor $N_c = 3$ arises from the trace over colours, $\mu$ is the renormalisation scale, while $q^\mu$ and $\varepsilon^\mu$ denote the momentum and the polarisation vector of the photon, respectively.  The Dirac trace evaluates to 
\beq \label{eq:A2}
{\rm tr} \left [   \sigma^{\mu \nu} \left ( \slashed{l} - \slashed {q} + m_t \right ) \slashed{\varepsilon}  \left ( \slashed{l} + m_t \right )  \right ] = 4 i \hspace{0.25mm} m_t \left ( q^\mu \varepsilon^\nu - q^{\nu} \varepsilon^\mu \right ) \,. 
\eeq
We hence find 
\beq \label{eq:A3}
{\cal A} = \frac{i}{4 \pi^4} \frac{3  \hspace{0.25mm}  e \hspace{0.25mm} Q_t }{M_\ast^2} \hspace{0.5mm} m_t \left (  \bar \chi \hspace{0.25mm}  \sigma_{\mu \nu} \hspace{0.25mm}  \chi \right) \left ( q^\mu \varepsilon^\nu  -  \varepsilon^\mu q^\nu \right ) \int \! d^d l \; \frac{ \mu^{4-d} }{\big ( \left ( l - q \right )^2 - m_t^2 \big) \left ( l ^2 - m_t^2 \right ) } \,. 
\eeq
In order to extract the contribution ${\cal A}_M$ to the magnetic dipole operator, we expand the integrand in powers of $q^2$ and obtain 
\beq \label{eq:A4}
{\cal A}_M = \frac{i}{4 \pi^4} \frac{3 \hspace{0.25mm}  e \hspace{0.5mm} Q_t}{M_\ast^2}  \hspace{0.5mm}  m_t \left (  \bar \chi \hspace{0.25mm}  \sigma_{\mu \nu} \hspace{0.25mm}  \chi \right) \left ( q^\mu \varepsilon^\nu  -  \varepsilon^\mu q^\nu \right ) \int \! d^d l \; \frac{ \mu^{4-d} }{ \left ( l ^2 - m_t^2 \right )^2 } \, .
\eeq
The loop integral in (\ref{eq:A4}) is readily evaluated as a power series in $ \epsilon = (4 - d)/2$. For energies too low to resolve the four-fermion tensor interactions, there is a UV $1/\epsilon$ pole. We regularise this singularity by the simple replacement 
\beq  \label{eq:A5}
 \frac{1}{\epsilon} + \ln \frac{\mu^2}{m^2} \to   \ln \frac{M_\ast^2}{m^2} \,.  
\eeq
After this replacement and keeping only the logarithmically enhanced terms, we arrive at 
\beq \label{eq:A6}
{\cal A}_M =  -\frac{3  \hspace{0.25mm} e \hspace{0.25mm} Q_t}{4 \pi^2} \frac{m_t}{M_\ast^2} \left (  \bar \chi \hspace{0.25mm}  \sigma_{\mu \nu} \hspace{0.25mm}  \chi \right) \left ( q^\mu \varepsilon^\nu  -  \varepsilon^\mu q^\nu \right )  \ln \frac{M_\ast^2}{m_t^2} \,.
\eeq
The result (\ref{eq:MLL}) is then obtained by replacing $\left ( q^\mu \varepsilon^\nu  -  \varepsilon^\mu q^\nu \right )$ with $F^{\mu \nu}$. The calculation of the ${\cal O}_{PT}$ insertion proceeds in exactly the same way as described above, and leads to an analogous result. 

We now turn to the computation of the two diagrams in Fig.~\ref{fig:diagramsloop}. In this case we can immediately neglect all external momenta. For the axialvector operator ${\cal O}_{AX}$ the sum of the two graphs gives
\beq \label{eq:A7}
\begin{split}
{\cal A} & = -\frac{1}{(2\pi)^4} \frac{1}{M_\ast^4} \, \mu^{4-d} \int \! d^d l \, \Bigg [ \frac{\left ( \bar \chi \hspace{0.25mm} \gamma_\mu \gamma_5 \hspace{-0.5mm} \left (\slashed{l} + m_\chi \right ) \hspace{-0.5mm} \gamma_\nu \gamma_5 \hspace{0.25mm} \chi \right ) \left ( \bar q \hspace{0.25mm} \gamma^\mu \gamma_5 \hspace{-0.5mm} \left (-\slashed{l} + m_q \right ) \hspace{-0.5mm} \gamma^\nu \gamma_5 \hspace{0.25mm} q \right )}{\left (l^2-m_\chi^2 \right ) \left (l^2-m_q^2 \right )} \\[1mm] 
& \hspace{4.5cm} + \frac{\left ( \bar \chi \hspace{0.25mm} \gamma_\mu \gamma_5 \hspace{-0.5mm} \left (\slashed{l} + m_\chi \right ) \hspace{-0.5mm} \gamma_\nu \gamma_5 \hspace{0.25mm} \chi \right ) \left ( \bar q \hspace{0.25mm} \gamma^\nu \gamma_5 \hspace{-0.5mm} \left (\slashed{l} + m_q \right ) \hspace{-0.5mm} \gamma^\mu \gamma_5 \hspace{0.25mm} q \right )}{\left (l^2-m_\chi^2 \right ) \left (l^2-m_q^2 \right )}  \Bigg ] \,.
\end{split}
\eeq
After some Dirac algebra, we obtain for the contribution ${\cal A}_S$ of this amplitude proportional to ${\cal O}_S$ the result
\beq  \label{eq:A8}
{\cal A}_S = -\frac{d}{8\pi^4}  \frac{ m_q \hspace{0.25mm} m_\chi}{M_\ast^4}  \left ( \bar \chi  \chi \right ) \left ( \bar q  q \right ) \int \! d^d l  \; \frac{\mu^{4-d} }{\left (l^2-m_\chi^2 \right ) \left (l^2-m_q^2 \right )}   \,.
\eeq
Expanding the resulting expression in $\epsilon$ and keeping only the UV pole, we find 
\beq  \label{eq:A9}
{\cal A}_S = -\frac{i}{2 \pi^2}  \frac{m_q \hspace{0.25mm} m_\chi}{M_\ast^4} \left ( \bar \chi  \chi \right )  \left ( \bar q q \right )   \hspace{0.25mm}  \frac{1}{\epsilon} \,.
\eeq
Identifying the singularity with the corresponding logarithm, we finally obtain (\ref{eq:axialloop}). In the case of ${\cal O}_{AN}$ the quark Dirac lines in (\ref{eq:A7}) do not contain $\gamma_5$ matrices, and in consequence the final result (\ref{eq:A9}) receives the opposite overall sign. 

%\bibliographystyle{JHEP}
%\bibliography{refs}

\providecommand{\href}[2]{#2}\begingroup\raggedright\endgroup

\end{document}